# Correlation and Network Topologies in Global and Local Stock Indices

Ashadun Nobi[1,2], Sungmin Lee[3], Doo Hwan Kim[4], and Jae Woo Lee[1]

[1]Department of Physics, Inha University, 100 Inha-ro, Nam-gu, Incheon 402-751 Korea

[2]Department of Computer Science and Telecommunication Engineering, Noakhali Science and Technology University, Sonapur Noakhali-3802, Bangladesh

[3]Department of Biochemistry and Molecular Biology, University of Southern Denmark, Campusvej 55, Odense 5230, Denmark

[4]Department of Physics, Incheon National University, 119 Academy-ro, Yeonsu-gu, Incheon 406-772 Korea

Abstract

This study examined how the correlation and network structure of 30 global indices and 145 local Korean indices belonging to the KOSPI 200 have changed during the 13-year period, 2000-2012. The correlations among the indices were calculated. The results showed that although the average correlations of the global indices increased with time, the local indices showed a decreasing trend except for drastic changes during crises. The average correlation of the local indices exceeded the global indices during the crises from 2000-2002, implying a strong correlation structure among the local indices during this period due to the detrimental effect of the 'dot-com bubble'. The threshold networks (TN) were constructed in the observation time window by assigning a threshold value and determining the network topologies. A significant change in the network topologies was observed due to the financial crises in both markets. The Jaccard similarities were also determined using the common links of TNs. The TNs of the financial network were not consistent with the evolution of the time, and the successive TNs of the global indices were more similar than those of the successive local indices. Finally, the Jaccard similarities identified the change in the market state due to a crisis in both markets.

1. Introduction

The growing interest of physicists in economic systems has led to the application of different techniques to the time series of financial markets [1-9]. The methods of correlation, random matrix theory (RMT), and network techniques have become important tools for analyzing financial data and extracting information on market movement, risk management, asset allocations, and the correlation structure of financial assets. Financial systems composed of wide variety of markets are positioned in different geographical locations respond differently to external information, such as the same economic announcements or market news [10-12]. This suggests that the experimental data of the financial time series carries economical information that is correlated. These types of global indices construct a global financial network and are reorganized due to a crisis or external information [13, 14]. The complex network has become an important tool for examining the properties of a complex system in different branches of science, such as social, biological, and financial systems [15-19].

On the other hand, financial systems composed of different types of companies are also responsive due to the same economic announcements and construct financial network [9]. The focus of this study was to observe and compare the change in the correlation and network properties of the global and local indices in the observation period. A sliding time window for one year was made. The aim was to analyze the long term effect, where all kinds of perturbations exist. Using the intra-day data, the correlation matrices were constructed and the average correlation was calculated. The network technique was then applied to the correlation matrices to construct the correlation networks. A recent study of threshold network of global indices was reported [13]. Here, the authors constructed a threshold network for different threshold values before and during a financial crisis and observed different types of network properties. The threshold network for local indices is reported elsewhere [9, 20]. These articles explain the degree distribution and reorganization of the local financial network due to global financial crises. In the present study, a threshold network (TN) of global and local market at a specific threshold for each year from 2000-2012 was constructed. Different kinds of network properties, such as the network density, characteristic path length, clustering coefficient for a global and local market, were determined and compared. The Jaccard Index or similarity was calculated to examine the market similarity and state [21]. In a recent article [22], the authors identified the market state by quantifying the difference in the correlation for two points in time. The Jaccard similarity (JS) assesses whether the subsequent occurrences of the network share many common links. These types of analyses help better understand the structural changes in the global and local financial network and enable the market state to be identified in the observation time window. The remainder of the paper is organized as follows: The financial data is discussed in section 2. Section 3 reports the results of correlation analysis. Section 4 discusses the method of network analysis and result. Section 5 reports the conclusions.

## 2. DATA ANALYZED

This study examined the daily closing prices of 30 stock market indices located all over the world and 145 indices from KOSPI 200 from 2000 to 2012. Appendix A shows the complete list of global indices. A list of the companies in Korean market is given in the supplementary information. The size of the time window was set to one year. In the analysis period, the market has been affected by different crises that reorganize the financial structures. Fig. 1 shows the 'dot-com' bubble in 2000 (A), September-11 attack in 2001(B), down turn of the stock prices in 2002 due to effects of the dot-com bubble and September-11 attack (C), subprime mortgage crisis in 2007(D), Global financial crisis 2008(E) and European sovereign debt (ESD) crisis 2011(F). The analysis was performed by the daily logarithmic return, which is defined for index *i* as follows:

$$R_i(t) = \ln[I_i(t)] - \ln[I_i(t-1)], \qquad (1)$$

where *I$_i$(t)* is the closing price of index *i* on day *t*. The average volatility for a particular period for a index is defined as

$$v(t) = \sum_{t=1}^{T-1} |R(t)|/(T-1). \qquad (2)$$

The normalized return for index *i* is defined as $r_i(t) = (R_i(t) - <R_i>)/\sigma_i$, where σ$_i$ is the standard deviation of the stock index time series *i*. The equal time cross-correlation at time *T* (approximately 260 days) was calculated from the normalized return of two indices by $C_{ij} = <r_i(t)\, r_j(t)>$.

## 3. Volatility and Correlation analysis

The wide swings in stock market prices all over the world in recent years have brought the financial community's interest in the concept of volatility. The stock market has shown considerable volatility all over the world in the recent past. Fig.1(a) plots the annual historical average volatility of some major stock market indices all over the world and KOSPI 200 local indices. The movement of volatility for both markets shows a similar trend. In 2000, the observed average volatility for the local market was extraordinarily higher than the global market. The average historical volatility for the local market in 2001 decreased by approximately, 26% whereas the decreased for the global market was approximately 9%. The effect of the IT bubble and the September-11 2001 attack can easily explain the high volatility in 2002 and 2001, in which the curve shows a plateau. After 2002, the volatility stabilized at a high level before resuming its upward trend in 2006 for the global market and 2007 for the local market.

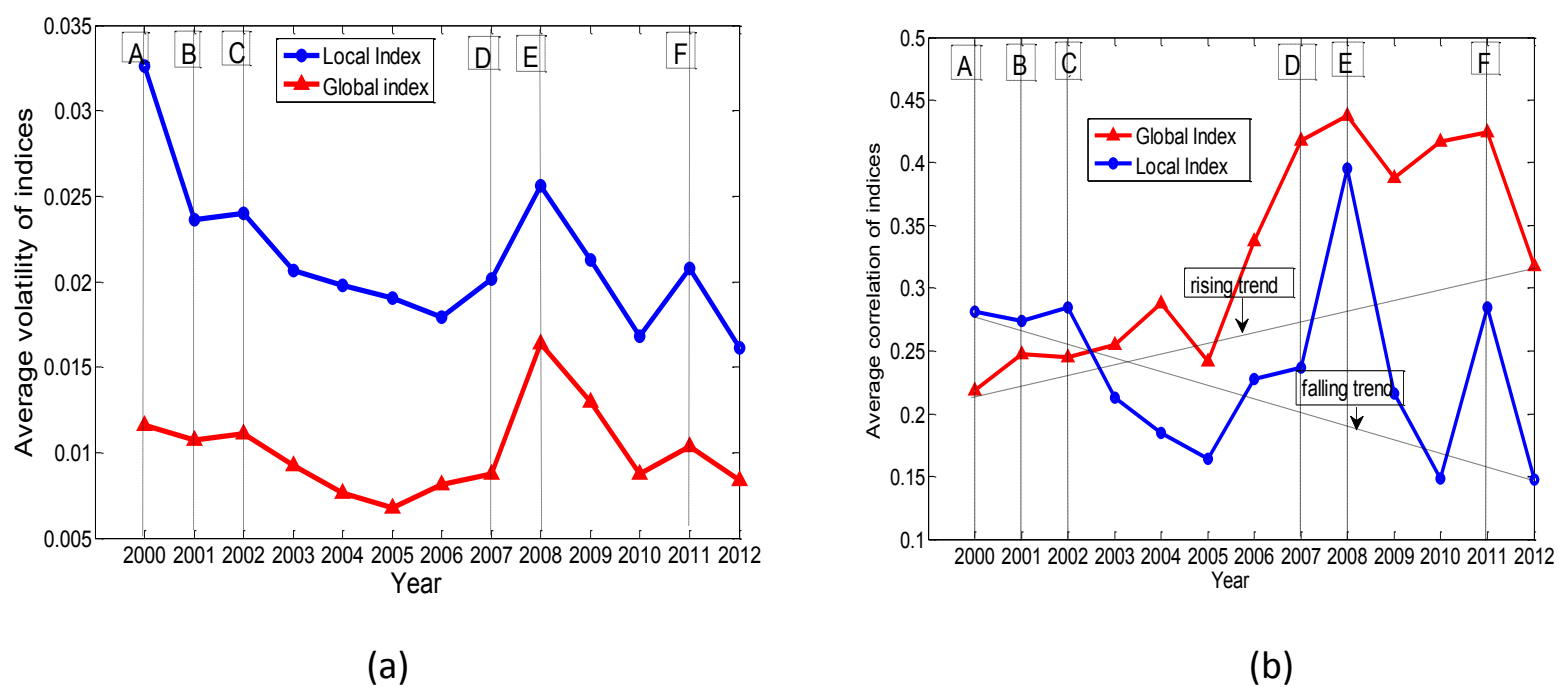

(a)  (b)

Figure 1. (a) Volatility trend of global and local indices (b) Average cross-correlation of the global and local indices.

The one year volatility indicator shows that the volatility reached higher levels (9.5% for Korean market and 22% for global market) due to the subprime crisis. During the global financial crisis in 2008, the volatility reached much higher levels (46% for global market and 20% for local market) than the level of volatility of the subprime crisis. The volatility then showed a downturn up to 2010 for the local and global markets, which was followed by an increasing trend again due to the European sovereign debt crisis in some states of Europe. The effect of this crisis caused an 18% increase in volatility for both markets than in 2010. The Pearson correlation coefficient was calculated to determine the degree of the correlation among the indices. Fig-1(b) shows the mean correlations of global and local indices. The average cross-correlations increased for the global market but decreased for the local market in the observation period except for sharp changes that occurred during the crises. The average correlation among the local indices is comparatively higher than the global market from 2000 to 2002 due to the dot-com bubble and September-11 2001 attack. The strong correlations among the local indices during the dot-com bubble indicate a drastic change in the market state due to the devastating effect of this crisis. On the other hand, the lower correlations among the global indices during the dot-com bubble indicate the lesser effect of this crisis in the global market. Subsequently, the average correlation of the local indices decreased (42%) from 2002 to 2005. whereas the correlation among the global indices increased almost 2%. A sharp change in the correlation was observed from 2006 due to the different crises for the global and local indices. The mortgage crisis in 2007, global financial crisis in 2008 and European sovereign debt crisis in 2011 make the correlation higher than the normal trend. After the global financial crisis, the correlation among the local indices decreased abruptly, whereas no sharp variation in the correlation was observed among the global indices. Over the last 13 years (2000-2012), the average correlation of the world indices increased approximately 31%. This increase in cross-regional equity correlation is a result of the integration of the global economies, global market, free trade between economies, the rise of the emerging market, and globalization of financial industry. [23]. On the other hand, the average correlation among the local indices over the last 13 years decreased by 41%. This may be due to a lack of integration among companies or strong grouping for similar kinds of companies. In addition, the strong correlation due to a crisis in the global market lasted longer than those for the local market.

4. Threshold networks

The construction of threshold networks (TN) assigning a certain threshold θ is familiar in financial networks [9, 20]. In the threshold network, a node (V) represents a distinct index and the links (E) represent the connection between two indices weighted by the cross-correlation between the two indices of a given year. In the threshold network, an undirected link adds nodes $i$ and $j$ if the correlation coefficient $C_{ij}$ is greater than or equal to $-1 \leq \theta \leq 1$. Therefore, the size of the formation of the cluster and the set of links of the cluster among the nodes depend on the value of the thresholds. The network properties of the largest cluster were constructed and analyzed. The constructed threshold networks were for a threshold of 0.3 for each year from 2000-2012. This threshold was selected because of the high rate of node connections to the largest cluster. The network properties of the largest cluster for the threshold $0.3 < \theta < 0.5$ (not shown) exhibited similar behavior to that with a threshold of 0.3. On the other hand, the networks that were fully connected at $\theta \leq 0.25$ and separated at different clusters for the thresholds around $\theta \geq 0.5$ [13] were not the focus of this study. Fig. 2 shows the financial threshold networks for the global and local indices for a threshold of 0.3. The number of connecting nodes to the largest cluster was 21 and 28 in 2000 and 2008 for the global indices, respectively. For the local market, however, the largest cluster sizes were 138 and 145 in 2000 and 2008, respectively. The network sizes of the largest cluster had changed due to the crises.

4.1 Network density

The ratio of the number of existing links to the maximum number of possible links is known as the network density, which can be determined as $\rho = M/[N(N-1)]$, where *N* is the total number of the nodes and *M* is the number of the connecting links [21]. Fig. 3(a) shows the network density of the largest clusters for the threshold networks at a threshold of 0.3. The network density for the global market fluctuated considerably up to 2005 and did not show any specific trend. This increased for the remainder of the study period with small fluctuations. The higher densities, where networks almost fully connected, were observed during the mortgage crisis (2007), global market crash (2008), and ESD crisis (2010 and 2011), which indicates a strong correlation among the indices during financial crises. In the local market, however, the densities were higher in 2000-2002 due to the effects of dot-com bubble and the September 11, 2001 attack, which then decreased up to 2005. Then the density increased, and after reaching the peak value during the stock market crash in 2008, it exhibited a decreasing trend and had high densities during the ESD crisis in 2011. When the global market was compared with the local market, the network densities of the global market were higher than those of the local market. This suggests that a global financial network is more correlated than a local one.

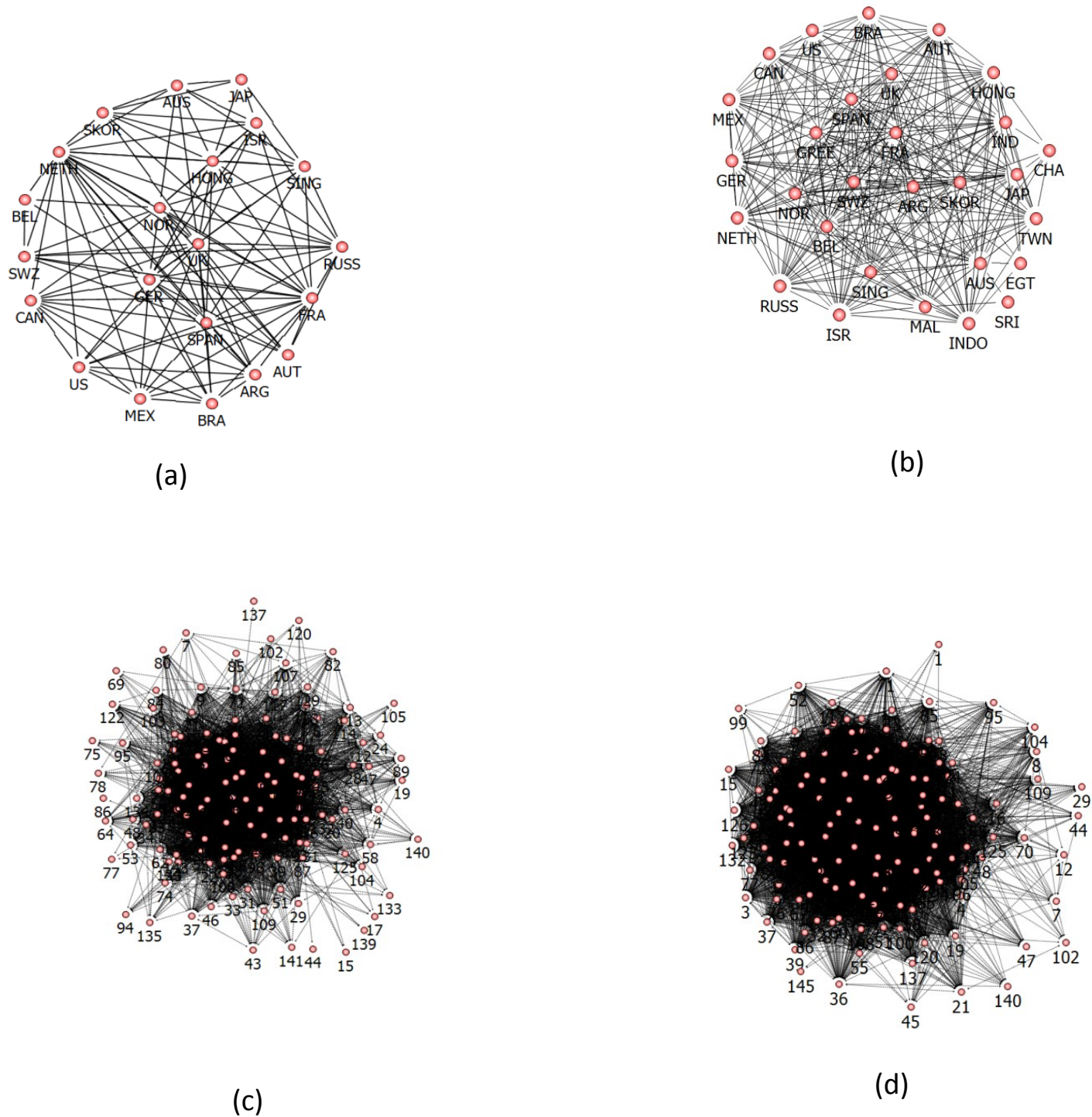

Fig. 2. Threshold networks at the threshold θ =0.3: (a) 2000, (b) 2008 for the global indices and (c) 2000, (d) 2008 for the local indices. Only the largest clusters are shown to avoid confusion.

4.2 Characteristic path length

The characteristic path length or the average shortest path length in a cluster can be expressed as

$$\bar{l} = \frac{2}{N(N-1)} \sum_{\substack{i,j \\ i<j}} l_{ij}, \qquad (3)$$

where $l_{ij}$ is the shortest path length between nodes $i$ and $j$. Fig.3 (b) shows the average shortest path length for the global and local market for the threshold networks at a threshold θ =0.3. The mean shortest path lengths for the global threshold networks from 2000 to 2005 were higher than in the other periods with significant fluctuations, which indicate weak correlations among the indices. On the other hand, during the different crises from 2006 to 2011, the average shortest path length became smaller, which indicates that a given node can be reached from the other nodes with a very small number of steps.

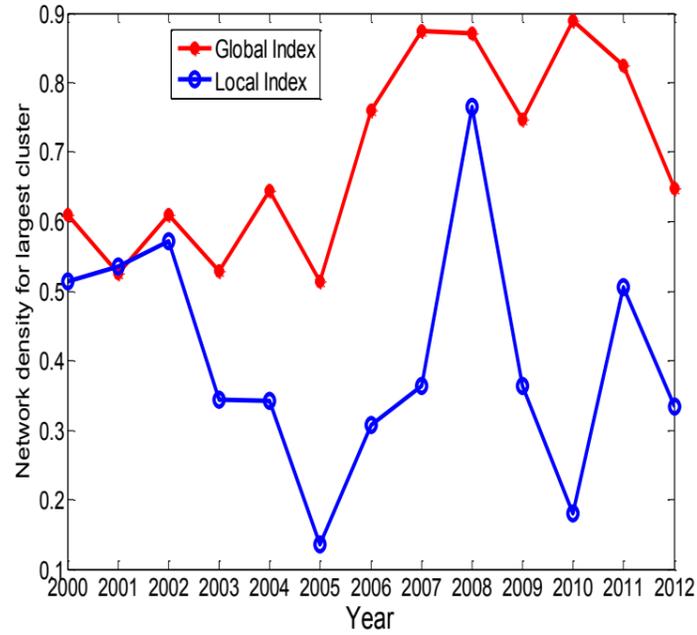

(a)

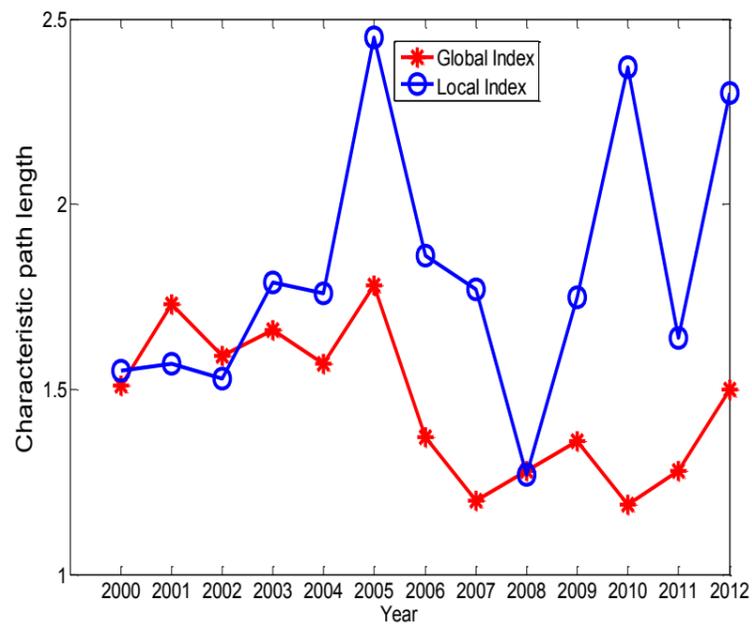

(b)

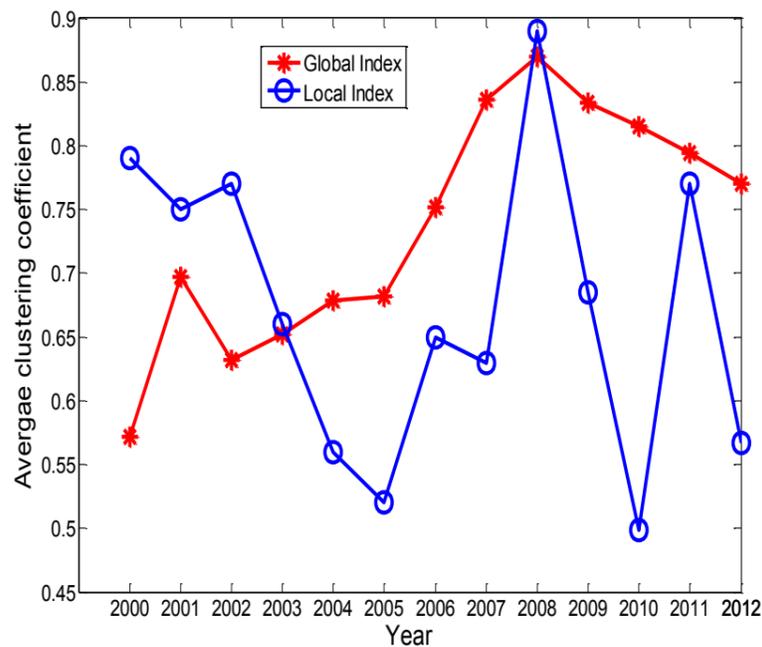

(c)

Fig.3. Topological properties of the largest cluster at a threshold of 0.3: (a) network density, (b) characteristic path length, and (c) average clustering coefficient.

This means that the indices come closer during a crisis. The lowest characteristic path lengths (≈1.2) were observed during the mortgage crisis (2007) and ESD crisis (2010). Again, the characteristic path length showed an increasing trend in 2012 when the crisis was overcome. In the local market, however, the trend of the average shortest path lengths was fluctuating with sharp peaks during the non-crisis periods. The average path length became smaller during the crises in 2000-2002 and the stock market crash in 2008. A sudden change in the average path length due to the crises was observed in the global and local markets.

4.3 Clustering coefficient

The average clustering coefficient is a measure of the local compactness of a network. The clustering coefficient of a vertex *i* can be expressed as

$$C_i = \frac{2m_i}{n_i(n_i-1)},\qquad(4)$$

where $n_i$ denotes the number of neighbors of vertex *i*, and $m_i$ is the number of the edges existing between the neighbors of vertex *i*. $C_i$ is equivalent to 0 if $n_i \leq 2$. The average clustering coefficient at a specific threshold for the entire network is defined as the average of $C_i$ over all the nodes in the network, i.e. $C = \frac{1}{N}\sum_{i=1}^{N} C_i$. Fig.3 (c) shows the average clustering coefficient of the global and local financial threshold networks at threshold θ =0.3. In the global market, the average clustering coefficient increases up to 2008 and after that, shows a decreasing trend. Higher values of the average clustering coefficient were observed during the different crises from 2006, suggesting that the networks are more clustered and compact during the crises. On the other hand, in the local market, the networks possess the high average clustering coefficient during the dot-com bubble and the september-11 2001 attack, and then decreased up to 2005. Subsequently, it showed an increasing trend up to the global financial market crash in 2008, where the highest average clustering coefficient was observed. In the remaining years, it showed a lower average clustering coefficient except for the ESD crisis in 2011. The average clustering coefficients in the local market showed high values during the different crises, such as the global market. The average clustering coefficient in the dot-com bubble (2000-2002) for the local market was much higher than those of the global market, indicating that the effect of the dot-com bubble in the local market was more destructive. Moreover, the global market showed a higher average clustering coefficient than the local market except for the large crises, suggesting that the global financial networks are more compact than local financial networks.

4.4 Jaccard similarity for threshold networks

The Jaccard similarities [21] among the threshold networks generated for each year at a specific threshold, θ =0.3, for global and local markets were calculated to observe the network similarity. The Jaccard similarity (JS) index of two MSTs can be defined as

$$J = \frac{N_1}{N-N_1}, \qquad (5)$$

where $N_1$ is the number of links with the same pair of nodes between the two TNs and $N$ is the total number of links of two TNs. This allows a comparison of the similarity and stability of the networks in the observation time window. Fig.4 shows a grayscale representation of the Jaccard similarities, in which each point is calculated from the largest cluster of the threshold networks over the 13 year period.

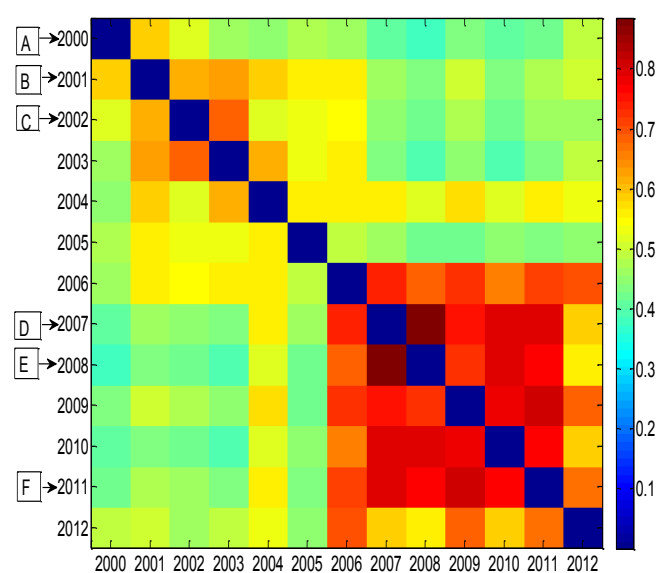

(a)

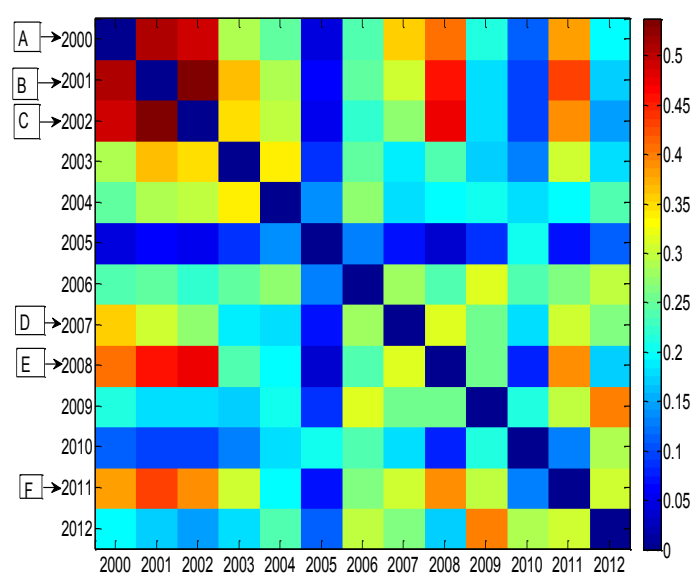

(b)

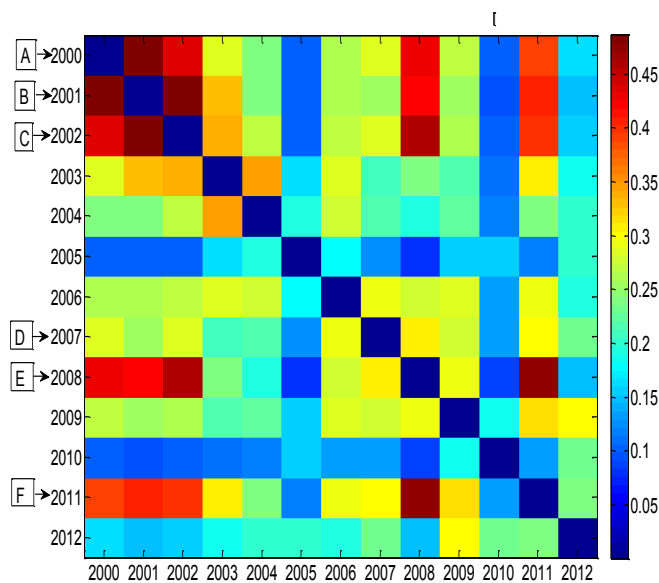

(c)

Fig.4. Grayscale representations of the time evolution of Jaccard similarity: (a) global indices, (b) 30 local indices, (c) 145 local indices. The change in color indicates a transition of the market.

Choose a point on the diagonal of this grayscale configuration and denote it as "now". From this point, the Jaccard similarity to the previous years can be found on the vertical line above this point, or the horizontal line to the left of this point. The market state can be identified easily from the changing JS. In the global market, the mean Jaccard similarity is 0.512, which indicates that two TNs can be similar at approximately 50%. On the other hand, the Jaccard similarity showed high values (more than 60%) from 2006 to 2011 when the mortgage crisis, the global financial crisis, and the European sovereign debt crisis occurred. The high Jaccard similarity (see red shaded areas in fig-4a) between two TNs suggests that the two networks or corresponding markets are under an external stress (crisis). The JS produced much lower values between the TN of the crisis and the non-crisis year, indicating a transition of the market state. In the local market, however, the average Jaccard similarity among TNs (approximately 25%) was not as high as the global market. The maximum Jaccard similarity for the local market was approximately 55% for the networks with 30 indices and 50% for the networks with 145 local indices, respectively, whereas it was approximately 80% for the global market. The states of the market from JS in the local market can be identified (red shaded areas in Fig. 4 (a) and Fig. 4 (c)). A comparison between two markets revealed the global market to have an approximately two times higher Jaccard similarity than the local market. This suggests that consecutive global networks are more similar than those of consecutive local networks. The reason of lower values of JS in the local market may be due to group dynamics. In addition, different market states were observed in the local market from the global market at events A-C due to severe effect of the 'dot-com' bubble. With the exception of events A-F in the global market, there was the mortgage crisis in 2006, meltdown of the crisis in 2009 and origin of the ESD crisis in 2010, which kept the global market in a similar state from 2006-2011. Based on the above discussions, the correlation networks and the state of financial markets was not consistent and the global market showed similar and dissimilar trends to those of the local market.

5. Conclusion

The correlation and network analysis of global and local financial markets were discussed and compared, particularly during crises. The average correlation of the global indices was always higher than the local indices except for the 'dot-com bubble', which suggests that the global indices are more correlated than the local indices. The higher average correlation of the local indices during the 'dot-com bubble' suggests that the effect of the bubble in the Korean market is more severe than the global market. The networks were constructed at a specific threshold for each year time window and the topological properties were examined. The topological properties of the financial correlation networks at a specific threshold showed an abrupt change in the presence of crisis events, highlighting the devastating effect of the 'dot-com bubble' on the Korean market. The overall changes in the TNs were examined using a common number of links. The Jaccard similarities suggest that the structure of the TN changes with time and the TNs of global financial market are more similar than those of the local market. Moreover, the change in the market state can be recognized easily with the change in JS. In conclusion, the empirical analysis provides evidence that the origin of the crisis, the effect of the crisis and the structural change in the correlation network due to the crisis in the local market are not entirely similar to those of the global market.

Acknowledgements

This study was supported by the research fund of the Science and Technology Policy Institute(STEPI).

Appendix: Global Stock Index

The following 30 world stock indices were considered. The European economic zone includes 11 countries as France (FRA), Germany (GER), the United Kingdom (UK), Espan (SPAN), Switzerland (SWIZ), Netherland (NETH), Belgium (BEL), Norway (NOR),Greece (GREE), Austria (AUT), and Russia (RUSS). In the Asian and Australian economic zone, 14 countries were included, such as Japan (JPN), South Korea (SKOR), Singapore (SING), Hong Kong (HONG), Indonesia (INDO), Egypt (EGT),Taiwan (TWN), Malaysia (MAL), China (CHA), India (IND), Pakistan (PAK), Srilonka (SRI), Israel (ISR), and Australia (AUS). The number of countries in the American economic zone was five: The United State of America (US), Canada (CAN), Mexico (MEX), Argentina (ARG), Brazil (BRA).